\documentclass[aps,pra,reprint,amsmath,amssymb]{revtex4-2}
\usepackage{bm}
\usepackage{physics}
\usepackage{mathtools}
\usepackage{amsthm}
\usepackage{xcolor}
\usepackage{comment}
\usepackage{appendix}
\usepackage[colorlinks=true, allcolors=blue]{hyperref}

\begin{document}

\title{Tennis-racket instability of twisted electrons}

\author{S.S. Baturin}
\affiliation{School of Physics and Engineering, ITMO University, St.\ Petersburg 197101, Russia}

\date{\today}

\begin{abstract}
We demonstrate that a weak nonlinear magnetic entrance edge induces a tennis-racket (Dzhanibekov) instability in the shell-resolved orbital pseudospin dynamics of twisted electrons propagating in a nominally uniform solenoidal field. Starting from a Maxwell-consistent thin-edge extension of the entrance field, we derive an effective fixed-shell Hamiltonian in which linear Schwinger pseudospin precession acquires an anisotropic quadratic correction. In the symmetric aligned limit, an exact linear eigenstate (a Laguerre–Gaussian vortex state) becomes a hyperbolic fixed point of the large-shell dynamics, producing recurrent reversals of the mean pseudospin projection. These reversals appear in real space as repeated conversions of the transverse profile between Laguerre–Gaussian vortex and Hermite–Gaussian multi-lobed states. The unavoidable Lewis–Ermakov breathing of realistic wave packets does not generate a separate mechanism; it naturally modulates the nonlinear strength and sets the growth time scale. Microscope-scale estimates show that the required regime is accessible with standard octupole correctors in a transmission electron microscope.
\end{abstract}

\maketitle

 Among the most striking instabilities of rotational motion is the tennis racket, or Dzhanibekov effect, where a rotation about the intermediate principal axis is unstable, and trajectories near the separatrix undergo recurrent flips (see for instance Ref.\cite{GeomTR2020}). Quantum counterparts have been discussed both as Bloch-sphere analogies in driven two-level systems \cite{Damme2017} and as genuine flipping dynamics of asymmetric quantum rotors \cite{TRNanoRot}. A more basic question remains: can an analogous instability appear in a purely wave-dynamical setting, without any mechanical rigid-body rotation? Twisted electrons in magnetic fields provide exactly such a setting.
Structured electron states in solenoidal fields support controlled transport and reshaping of orbital modes \cite{Exp0,Exp01,Exp02,Exp1,Exp2,Halfring,modeconv}. When the axial symmetry of the entrance field is broken, the transverse post-edge dynamics reduces to an exact two-dimensional problem whose fixed-shell sectors carry a Schwinger $  \mathfrak{su}(2)  $ algebra or equivalently, an $  SO(3)  $ dynamics on the orbital Poincar\'e sphere \cite{Filina2024}. This description is natural in the geometric-phase theory of structured waves, where effective two-state sectors are represented on Poincar\'e-type spheres by pseudospin variables rather than literal real space angular-momentum operators \cite{VANENK1993,Padgett:99,Galvez2003,Calvo:05,RevModPhys2022}. We therefore call the shell-resolved degree of freedom an orbital pseudospin, an internal mode variable acting within a degenerate Landau-level shell, not a directly measurable orbital angular momentum vector. Only in the symmetric aligned limit is one component proportional to the canonical operator $  L_z  $ \cite{Bliokh2012}.
On a fixed shell, the poles of the orbital Poincar\'e sphere correspond to Laguerre–Gaussian vortex states, while the equatorial belt corresponds to Hermite–Gaussian states, the direct analogue of the familiar first-order structured light sphere, now realized for an electron wavepacket. A reversal of the mean orbital pseudospin projection therefore appears not as a change in measured angular momentum, but as a characteristic mode conversion between vortex-like LG and multi-lobed HG profiles, accompanied by a global restructuring of the wavepacket phase.
In the linear theory, the magnetic edge fixes the post-entry transport generator and induces only rigid precession of the orbital pseudospin about an axis set by the entrance geometry \cite{Filina2024}. Linear precession alone cannot produce a hyperbolic fixed point, a separatrix, or recurrent reversals. Because the edge field enters the gauge-fixed post-edge Hamiltonian, however, a weak nonlinear magnetic boundary can imprint a persistent anisotropic quadratic correction. In this paper we show that a Maxwell-consistent thin-edge extension of the solenoidal field does precisely that. In the symmetric aligned limit an LG state that is an exact linear eigenstate becomes a hyperbolic fixed point of the large-shell dynamics. The resulting motion is the direct wave-dynamical analogue of the Dzhanibekov instability where trajectories near the separatrix exhibit recurrent reversals of the mean orbital pseudospin projection, manifested in real space as repeated LG to HG mode conversion. We further demonstrate that the unavoidable Lewis–Ermakov breathing \cite{VanEnk2021,Silenko2021,Filina2023,Filina2026} of realistic wavepackets does not introduce a separate mechanism. Instead it naturally modulates the nonlinear strength and sets the instability onset scale. Numerical trajectories display clean recurrent flips, and microscope-scale estimates show that the required regime is accessible with standard octupole correctors. Twisted-electron transport in a nominally uniform magnetic field thus provides a concrete, controllable wave-dynamical realization of the Dzhanibekov effect on an orbital Poincar\'e sphere. 

Throughout the paper we use natural units $\hbar=c=1$.

We start from the thin-edge solenoidal model introduced in Ref.~\cite{Filina2024}.
Consider the step-like longitudinal field
\begin{equation}
B_{\tilde {z}}=\tilde{\Delta}_\perp \tilde\Phi_2 \theta(\tilde z),
\label{eq:Bz}
\end{equation}
supplemented by the divergence-free transverse discontinuity that models the transverse field components at the entrance of the coil,
\begin{equation}
B_{\tilde{x}}=-\partial_{\tilde x}\tilde\Phi_2 \delta(\tilde z),\qquad
B_{\tilde y}=-\partial_{\tilde y} \tilde\Phi_2 \delta(\tilde z),
\label{eq:Blin}
\end{equation}
where we have introduced the quadratic scalar potential function $\tilde \Phi_2$ of the form 
\begin{align}
    \tilde \Phi_2=B_0\left[\frac{(1-\beta)}{2}\tilde x^2+\frac{\beta}{2}\tilde y^2\right]
\end{align}
and \(\beta\in[0,1]\) is the symmetry parameter.
The corresponding near-axis vector potential can be gauge-fixed so that only the transverse components depend on the transverse coordinates $\tilde{\bm A}^T = (-\partial_{\tilde{y}} \tilde{\Phi}_2 \theta(\tilde{z}),\partial_{\tilde{x}} \tilde{\Phi}_2\theta(\tilde{z}),0)$. The explicit form of the transverse part reads
\begin{equation}
\tilde {\bm A}_\perp^T
=
B_0\left(-\beta\theta(\tilde z)\tilde y,\,(1-\beta)\theta(\tilde z) \tilde x\right),
\qquad
\tilde A_z=0.
\label{eq:Apra}
\end{equation}
This gauge is in Coulomb form, \(\tilde \nabla\cdot \tilde{\bm A}=0\) and yields a strictly two-dimensional Schr\"{o}dinger-like transverse setup in the Foldy-Wouthuysen representation under the paraxial reduction ($k_\perp \ll k_z$) with spin neglected \cite{SilenkoFW,SilenkoG,Silenko2021,Filina2026}. The natural scales of the problem are $k\rho_H^2$ for the longitudinal coordinate $\tilde z$ and
$\rho_H=\sqrt{\frac{2}{|e|B_0}}$ for the transverse coordinates $\tilde x,\tilde y$.
Here $k$ is the longitudinal wave vector defined by $k^2=E^2-m^2$, where $E$ is the total energy. We now switch to normalized coordinates $x=\tilde x/\rho_H$, $y=\tilde y/\rho_H$, $z=\tilde z/(k\rho_H^2)$ and consider the paraxial reduction of the Dirac equation with spin neglected in the form
\begin{equation}
i\partial_z \psi
=
\hat{H}_\perp \psi, ~~ \hat{H}_\perp=\frac{[\hat{\bm p}_\perp-\operatorname{sgn}(e)\bm A_\perp]^2}{2}.
\label{eq:Hperp}
\end{equation}
With the canonical rescaling
\begin{align}
&\bar x=\sqrt{2(1-\beta)}\,x,\qquad
\bar y=\sqrt{2\beta}\,y,\nonumber\\
&\bar p_x=\frac{p_x}{\sqrt{2(1-\beta)}},\qquad
\bar p_y=\frac{p_y}{\sqrt{2\beta}},
\label{eq:rescaling}
\end{align}
the Hamiltonian assumes the Schwinger form\cite{Schwinger,Agarwal_2006,Chen2011,Filina2024}
\begin{align}
&\hat{\bar{H}}_\perp=\hat{\bar{H}}_s+\hat{\bar{H}}_{as},\nonumber\\
&\hat{\bar{H}}_s=\frac{\hat{\bar{p}}^2_\perp}{2}+\frac{\hat{\bar \rho}^2}{2},\nonumber\\
&\hat{\bar{H}}_{as}(\alpha)=-\operatorname{sgn}(e)\bigl[\cos(2\alpha)\hat{\bar{H}}_1+\sin(2\alpha)\hat{\bar{H}}_3\bigr],
\label{eq:HsHas}
\end{align}
where $\sin^2(\alpha)=\beta$, $\hat{\bar {\rho}}^2=\hat{\bar {x}}^2+\hat{\bar {y}}^2$ and
\begin{equation}
\hat{\bar{\mathcal L}}_i=\frac{\hat{\bar{H}}_i}{2},
\qquad
[\hat{\bar{\mathcal L}}_i,\hat{\bar{\mathcal L}}_j]=i\epsilon_{ijk}\hat{\bar{\mathcal L}}_k.
\label{eq:Li}
\end{equation}
Here $\hat{\bar{\mathcal{L}}}_i$ are operators of orbital pseudospin projections or the generators of rotation about three orthogonal axes on the orbital Poincar\'e sphere \cite{RevModPhys2022,Filina2024}. It is known that $\hat{\bar{\mathcal{K}}}=\hat{\bar{H}}_s^2/4-1/4$ is a Casimir operator of the $\hat{\bar{\mathcal{L}}}_i$ algebra, thus for the on-shell state (the state that belongs to a specific orbital Poincar\'e sphere i.e. has fixed $j=n_r+|l|/2$ or fixed transverse energy) the Hamiltonian reduces to the rotation about the axis on the orbital Poincar\'e sphere that is set by the symmetry of the transverse magnetic field of the solenoid.

Namely, the shell Hamiltonian is given by 
\begin{equation}
\hat{\bar{H}}_{\mathrm{shell}}
=
-2\operatorname{sgn}(e)\bm n\cdot \hat{\bar{\bm{\mathcal L}}},
\label{eq:Hshell}
\end{equation}
and any incoming same-shell state that is not an eigenstate of \(\hat H_{\mathrm{shell}}\) undergoes rigid precession about the axis $n$ that is defined by both the asymmetry of the magnetic coil and the rotation of the coil about the $z$ axis with respect to the laboratory frame \cite{Filina2024}.

For clarity, however, we now specialize to the \emph{fully symmetric aligned case}
\begin{equation}
\alpha=\frac{\pi}{4},
\qquad
\beta=\frac12,
\label{eq:aligned}
\end{equation}
for which the canonical rescaling is trivial and a physical Laguerre-Gaussian (LG) mode is an exact eigenstate of the linear problem. Hence we drop bars in our further calculations. In that case the shell Hamiltonian for the electron $e<0$ reduces to
\begin{equation}
\hat H_{\mathrm{shell}}=2\hat{\mathcal L}_3.
\label{eq:Hshell_symmetric}
\end{equation}

A small mismatch between the incoming mode axis and the Hamiltonian axis can be represented equivalently either as a slight asymmetry of the entrance field or as a small rotation of the incoming same-shell state. For clarity we adopt the latter representation and keep the body Hamiltonian fully symmetric.
Thus we take the incoming state to be a slightly rotated extremal LG state,
\begin{equation}
\ket{\psi_{\mathrm{in}}}
=
e^{-i\delta \hat{\mathcal L}_2}\ket{j,j},
\qquad
0<\delta\ll 1.
\label{eq:seededstate}
\end{equation}
Expanding to first order in \(\delta\),
\begin{equation}
\ket{\psi_{\mathrm{in}}}
=
\ket{j,j}
-\frac{\delta}{2}\sqrt{2j}\,\ket{j,j-1}
+O(\delta^2).
\label{eq:seededstate_expanded}
\end{equation}
This keeps the state on the same orbital Poincar\'e sphere while introducing a controlled transverse seed.

The corresponding first moments for the case of $j\gg 1$ are
\begin{equation}
\ev{\hat{\mathcal L}_1}\simeq j\delta,~~
\ev{\hat{\mathcal L}_2}=0,~~
\ev{\hat{\mathcal L}_3}\simeq j\qty(1-\frac{\delta^2}{2}),
\label{eq:firstmoments_seed}
\end{equation}
so the deterministic transverse seed is of order \(j\delta\).
At the same time, the transverse fluctuations near the extremal shell state scale as
\begin{equation}
\Delta \mathcal L_1\sim \Delta \mathcal L_2\sim \sqrt{\frac{j}{2}},
\label{eq:fluct_seed}
\end{equation}
hence the intrinsic angular width on the orbital Poincar\'e sphere scales as \(j^{-1/2}\).
Therefore the quasiclassical first-moment description is reliable in the window
\begin{equation}
j^{-1/2}\ll \delta \ll 1.
\label{eq:delta_window}
\end{equation}
We emphasize that the quasiclassical closure is \emph{not} invoked for the exact pole state \(\ket{j,j}\) itself; the slightly rotated state \eqref{eq:seededstate} provides the seed that triggers the precession needed further. For experimentally accessible \cite{Exp02,Exp2,ElectronVort} topological charges $  |l| = 200  $–$  2000  $ ($  j = 100  $–$  1000  $) the window $  j^{-1/2} \ll \delta \ll 1  $ is comfortably satisfied by $  \delta \approx 0.1  $–$  0.2  $, a value readily achieved by a slight intentional misalignment of the incoming mode or by residual imperfections of the holographic phase plate.

The linear model above is purely quadratic and therefore generates only rigid precession. To go beyond that, we introduce a nonlinear thin-edge octupole perturbation. This perturbation preserves the two-dimensional structure of the post-edge Hamiltonian while enabling substantially richer dynamics.

With the next step we add a thin normal octupole magnet at the edge that is routinely used as a corrector in an electron microscope. This can be accomplished in the model through the quartic potential function that is harmonic in the transverse plane,
\begin{equation}
\Phi=\Phi_2+\Phi_4^{(n)},
\label{eq:Phifull}
\end{equation}
with
\begin{equation}
\Phi_4^{(n)}=\varkappa(x^4-6x^2y^2+y^4),~~\varkappa=\frac{\kappa_n \rho_H^2}{2 B_0 R_c^2}
\label{eq:quartics}
\end{equation}
and $R_c$ being the bore radius of the octupole and $\kappa_n=B_{\mathrm {tip}} L/R_c$ with $B_{\mathrm{tip}}$ the magnetic field at the pole tip and $L$ is the length (thickness) of the octupole.
The normalized symmetric solenoid edge potential function is 
\begin{align}
    \Phi_2=\frac{x^2+y^2}{2}
\end{align}
Because \(\Phi_4^{(n)}\) is harmonic, the bulk field (inside the solenoid) remains unchanged, while the transverse edge kick acquires cubic corrections.

We assume $\varkappa$ to be small. Thus, to first order in $\varkappa$, the quartic edge induces a quartic perturbation of the transverse Hamiltonian and a cross term 
\begin{equation}
\hat H_\perp
=
\hat H_{\mathrm{s}}+2\hat{\mathcal L}_3+\nabla_\perp \Phi_4^{(s)} \cdot\hat{\bm p}_\perp
+\hat V_4
+O(\varkappa^2).
\label{eq:Hquartic}
\end{equation}
Above $\Phi_4^{(s)}$ stands for the skew octupole potential function given by
\begin{align}
    \Phi_4^{(s)}=4\varkappa(x^3y-xy^3)
\end{align}
and we used the fact that skew and normal octupole potential functions are harmonic conjugates $\partial_x \Phi_4^{(s)}=-\partial_y \Phi_4^{(n)}$ and $\partial_y \Phi_4^{(s)}= \partial_x \Phi_4^{(n)}$. We note that the resulting gradient $\nabla_\perp \Phi_4^{(s)}$ is understood as an operator. 
Quartic potential \(\hat V_4\) is given by (see details in the Appendix \ref{app:field})
\begin{align}
    \hat{V}_4^{(n)}=\frac{2\kappa_n \rho_H^2}{B_0 R_c^2}\left[\hat{x}^4-6\hat{x}^2\hat{y}^2+\hat{y}^4\right].
\end{align}
Within one fixed shell \(j\), the projected on-shell perturbation reduces to a scalar and a quadratic orbital pseudospin form, the cross term vanishes exactly
\begin{align}
&\hat{P}_j\hat V_4 \hat{P}_j
=
c_0+\mu_0(\hat{\mathcal L}_1^2-\hat{\mathcal L}_2^2), \\
&\hat{P}_j\nabla_\perp \Phi_4^{(s)} \cdot\hat{\bm p}_\perp \hat{P}_j=0. \nonumber
\label{eq:Vnormal}
\end{align}
with $\mu_0=24\varkappa =12\frac{\kappa_n \rho_H^2}{B_0 R_c^2}$ being the static normalized strength parameter.
Above the shell projector is  $\hat{P}_j=\sum\limits_{m=-j}^{j} |j,m\rangle \langle j,m|$ with the $j$ fixed.
This reduction is derived explicitly in the Appendix \ref{app:projection}.

We therefore study the effective shell Hamiltonian
\begin{equation}
\hat H_{\mathrm{eff}}^{(j)}
=
2\hat{\mathcal L}_3+\mu_0(\hat{\mathcal L}_1^2-\hat{\mathcal L}_2^2).
\label{eq:targetH}
\end{equation}

Using the commutator relations \eqref{eq:Li}, the exact Heisenberg equations are
\begin{align}
\dot{\hat{\mathcal L}}_1
&=
 -2\hat{\mathcal L}_2
-\mu_0\{\hat{\mathcal L}_2,\hat{\mathcal L}_3\},
\nonumber
\\
\dot{\hat{\mathcal L}}_2
&=
2\hat{\mathcal L}_1
-\mu_0\{\hat{\mathcal L}_1,\hat{\mathcal L}_3\},
\label{eq:HL1}
\\
\dot{\hat{\mathcal L}}_3
&=
2\mu_0\{\hat{\mathcal L}_1,\hat{\mathcal L}_2\}.
\nonumber
\end{align}
Here and further the dot above the letter denotes the total derivative with respect to $z$ i.e. $\dot{\hat{\mathcal L}}_i\equiv \frac{d{\hat{\mathcal L}}_i}{dz}$. Equations \eqref{eq:HL1} are exact on each fixed Schwinger shell.

We introduce the normalized expectation values
\begin{equation}
\ell_i=\frac{\ev{\hat{\mathcal L}_i}}{j}.
\label{eq:elli}
\end{equation}
Taking expectation values of Eqs.~\eqref{eq:HL1} gives
\begin{align}
\dot \ell_1
&=
 -2\ell_2
-\frac{\mu_0}{j}\ev{\{\hat{\mathcal L}_2,\hat{\mathcal L}_3\}},
\nonumber
\\
\dot \ell_2
&=
2\ell_1
-\frac{\mu_0}{j}\ev{\{\hat{\mathcal L}_1,\hat{\mathcal L}_3\}},
\label{eq:ell1_exact}
\\
\dot \ell_3
&=
\frac{2\mu_0}{j}\ev{\{\hat{\mathcal L}_1,\hat{\mathcal L}_2\}}.
\nonumber
\end{align}
At this stage the dynamics is not closed, since it depends on second moments.

To obtain a closed quasiclassical dynamics, we assume that the state remains narrowly localized on a fixed shell \(j\), so that connected second cumulants are subleading
\begin{equation}
\ev{\{\hat{\mathcal L}_i,\hat{\mathcal L}_j\}}
=
2\,\ev{\hat{\mathcal L}_i}\,\ev{\hat{\mathcal L}_j}
+O(j),
\label{eq:factorization}
\end{equation}
with $\ev{\hat{\mathcal L}_i}=O(j)$. Under the large $j$ assumption $j^{-1/2}\ll 1$, Eqs.~\eqref{eq:ell1_exact} reduce to
\begin{align}
\dot \ell_1
&=
 -2\ell_2
-2\mu_0 j\,\ell_2\ell_3,
\nonumber
\\
\dot \ell_2
&=
2\ell_1
-2\mu_0 j\,\ell_1\ell_3,
\label{eq:QuasiCLsys1}
\\
\dot \ell_3
&=
4\mu_0 j\,\ell_1\ell_2.
\nonumber
\end{align}
These closed equations preserve
\[
\ell_1^2+\ell_2^2+\ell_3^2=1
\]
and define the quasiclassical flow on the orbital Poincar\'e sphere.

%+++++++++++++++++++++++
\begin{figure*}[t]
    \centering
\includegraphics[width=1\textwidth]{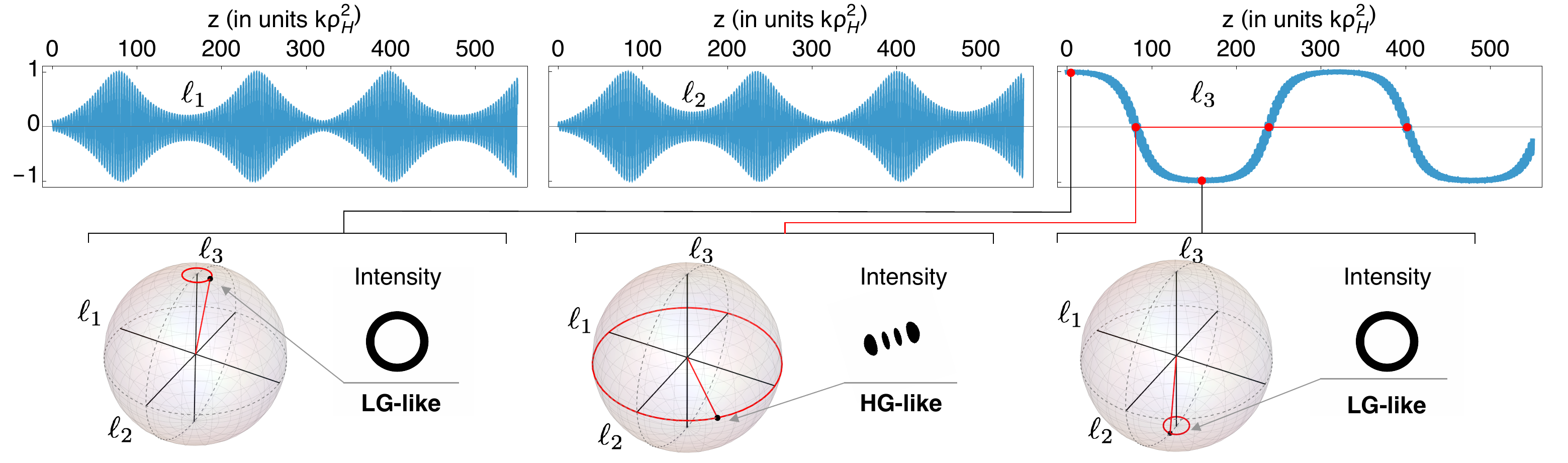}
\caption{Upper panels: numerical solution of Eqs. \eqref{eq:ressys} showing the switch-like evolution of the shell mean orbital pseudospin projections $\ell_i$ for the parameters mentioned in the text and initial state mismatch of $\delta=0.1$. Lower panels: schematic orbital Poincar\'e sphere interpretation and corresponding transverse intensity profile evolution for a narrow fixed-$j$ wave packet. As $\ell_3$ crosses through zero, the profile switches from LG-like to HG-like and back to LG-like. The lower intensity sketches are shown for reduced initial OAM for visual clarity.}
%+++++++++++++++++++++++
\label{fig:Fig2}
%+++++++++++++++++++++++
\end{figure*}
%+++++++++++++++++++++++

Introducing the control parameter
\begin{equation}
\mu=\frac{j\mu_0}{2} ,
\label{eq:mu_again}
\end{equation}
one checks that Eqs.~\eqref{eq:QuasiCLsys1} are generated by the dimensionless classical energy
\begin{equation}
\mathcal E(\ell_1,\ell_2,\ell_3)
=
2\ell_3+2\mu(\ell_1^2-\ell_2^2).
\label{eq:Eclassical_from_heis}
\end{equation}
Equation \eqref{eq:Eclassical_from_heis} is therefore the large-\(j\) reduction of the operator dynamics.

We linearize Eqs.~\eqref{eq:QuasiCLsys1} near the positive-\(\ell_3\) pole,
\begin{equation}
\ell_1,\ell_2\ll 1,\qquad \ell_3\simeq 1.
\label{eq:north_lin}
\end{equation}
Then
\begin{align}
\label{eq:lin_north1}
\dot \ell_1 = -(4\mu+2)\ell_2, ~~\dot \ell_2 = -(4\mu-2)\ell_1,
\end{align}
and
\begin{equation}
\ddot \ell_{1,2}
=
(16\mu^2-4)\ell_{1,2}.
\label{eq:lin_instability}
\end{equation}
Hence the aligned LG pole becomes hyperbolically unstable at
\begin{equation}
\mu>\frac12.
\label{eq:instability_again}
\end{equation}
This is the onset of the unstable-axis regime.

The separatrix reaches the opposite-\(\ell_3\) hemisphere if
\begin{equation}
\mu=\frac{\mu_0 j}{2}>1.
\label{eq:flipthreshold}
\end{equation}
Indeed, the separatrix energy is the saddle energy,
\begin{equation}
\mathcal E_{\mathrm{sep}}=2.
\end{equation}
Solving
\begin{equation}
2\ell_3+2\mu(\ell_1^2-\ell_2^2)=2,
\qquad
\ell_1^2+\ell_2^2+\ell_3^2=1,
\end{equation}
one finds that the minimum \(\ell_3\) reached on the separatrix is
\begin{equation}
\ell_{3}^{\mathrm{min}}=\frac{1}{\mu}-1.
\label{eq:u3min}
\end{equation}
Therefore the separatrix reaches $\ell_3<0$ iff $\mu>1$.
This is the direct analogue of the Dzhanibekov instability in the reduced orbital pseudospin dynamics.
While this simplified model already establishes the existence of the instability at the conceptual level,
a brief estimate of the parameter $\mu_0$ shows that, for a reasonable octupole strength, the instability
is feasible in the static model, although it may be challenging to observe directly.
Indeed, if we take $j=1000$, $B_{\mathrm{tip}}=1~\mathrm{T}$, solenoidal field $B_0=8\times10^{-3}~\mathrm{T}$,
octupole bore radius $R_c=1~\mathrm{mm}$, and octupole length $L=1~\mathrm{cm}$, we obtain
$\mu_0\approx 2.5\times10^{-3}$ and $\mu\approx 1.2$, which is above both the instability threshold Eq.\eqref{eq:instability_again} and the flip threshold Eq.\eqref{eq:flipthreshold}. For a $100~\mathrm{keV}$ electron, the characteristic Landau orbit size is $\rho_H\approx 0.41~\mu\mathrm{m}$, which is reasonable in terms of the beam size. We note that the small nonlinear perturbation regime holds at the Hamiltonian level as the strength parameter $\mu_0$ is well below unity. At the same time, the large shell excitation $j$ plays the role of the amplifier and opens an instability channel by pushing the control parameter $\mu$ above one half.  

To relax the conditions on the octupole strength and to demonstrate greater experimental flexibility, we take a step further and analyze a more realistic dynamical scenario where this instability is driven by the well-known breathing effect \cite{Silenko2021,VanEnk2021,Filina2023}.

We pick a breathing Lewis basis discussed in Refs.\cite{Filina2023,Filina2026,EpovArxiv}. In this case the Schwinger generators are defined by the Lewis ladder operators \cite{Lewis,Lewis1}, and therefore satisfy the same algebra on each fixed shell (see Appendix \ref{app:LEquartic}). Under the Ermakov \cite{Filina2023,Aldaya_2011,Guerrero_2013,Guerrero_2014} scaling \(x=b(z)x^E\), \(y=b(z)y^E\), the normal octupole quartic acquires the factor $b^4(z)$. Hence the projected shell Hamiltonian becomes
\begin{equation}
\hat H_{\rm eff}^{E}(z)
=
2\hat{\mathcal L}^E_3(z)
+
\mu_0 b^4(z)\left[(\hat{\mathcal L}^E_1(z))^2-(\hat{\mathcal L}^E_2(z))^2 \right],
\end{equation}
where as before $\mu_0= 12\frac{B_{\mathrm{tip}} \rho_H^2}{B_0 R_c^2}\frac{L}{R_c}$ is the static strength parameter.
The exact Heisenberg equations are therefore
\begin{align}
\dot{\hat{\mathcal L}}^E_1
&=
-2\hat{\mathcal L}_2^E
-
\mu_0 b^4(z)\{\hat{\mathcal L}^E_2,\hat{\mathcal L}^E_3\},\nonumber
\\
\dot{\hat{\mathcal L}}^E_2
&=
2\hat{\mathcal L}_1^E
-
\mu_0 b^4(z)\{\hat{\mathcal L}_1^E,\hat{\mathcal L}_3^E\},
\\
\dot{\hat{\mathcal L}}_3^E
&=
2\mu_0 b^4(z)\{\hat{\mathcal L}_1^E,\hat{\mathcal L}_2^E\}.\nonumber
\end{align}
For a narrow large-\(j\) packet this yields the quasiclassical system that is completely analogous to the static system \eqref{eq:QuasiCLsys1}
\begin{align}
\dot{\ell}_1^E
&=-
2\ell_2^E-4\mu b^4(z)\ell_2^E\ell_3^E, \nonumber
\\
\dot{\ell}_2^E
&=
2\ell_1^E-4\mu b^4(z)\ell_1^E\ell_3^E,
\label{eq:ressys}
\\
\dot{\ell}_3^E
&=
8\mu b^4(z)\ell_1^E\ell_2^E. \nonumber
\end{align}
The breathing envelope enters explicitly as a parametric modulation of the nonlinear top. We show in Appendix \ref{app:resonance} that, beyond the general multiplicative enhancement due to the large maximum of $b(z)$,
the breathing is exactly resonant and therefore drives the instability exponentially, further relaxing the condition on the magnitude of $\mu$. The Floquet exponent for large mismatch between the incoming state size and the Landau radius scales as $\propto \mu b_0^4$,  making the instability observable even when the static $\mu$ is far below the threshold. In Fig.\ref{fig:Fig2} we plot numerical solution to the system \eqref{eq:ressys} (with $b(z)$ given in Appendix \ref{app:ermakovsol}) for the set of experimentally reasonable parameters. We take solenoidal field of $B_0=0.5~\mathrm{T}$ and octupole tip field  of $B_{\mathrm{tip}}=0.1~\mathrm{T}$, octupole bore radius $R_c=6~\mathrm{mm}$, octupole length $L=1~\mathrm{cm}$ and mismatch between the Landau radius $\rho_H=51~\mathrm{nm}$ and state size at the entrance $\sim 1.7~\mu\mathrm{m}$ that correspond to the initial condition for the breathing radius $b_0=33$ and $b'_0=0$. State mismatch parameter is taken $\delta=0.1$ according to the prescription Eq.\eqref{eq:delta_window}. For $j=1000$ this results in the control parameter $\mu=1.5\times10^{-7}$. Despite the small value of the static control parameter the instability still develops due to the resonance at the distance of $\approx 82k\rho_H^2$ (as can be seen from Fig.\ref{fig:Fig2}). For the $100~\mathrm{keV}$ electron this is $37~\mathrm{cm}$, a distance that is reasonable for a microscope-based experiment. We note that lowering the energy to $10~\mathrm{keV}$ will result in even smaller  ``flip'' distance of $11~\mathrm{cm}$.

We emphasize that the orbital pseudospin description we provide should be understood as an effective reduced-shell representation of the dynamics rather than as an independently measurable internal degree of freedom. 
Nevertheless, in the high-$j$, well-localized regime, the corresponding motion on the shell sphere has a direct optical signature, namely, the instability-driven orbital pseudospin flip is accompanied by a pronounced deformation of the transverse intensity profile, approximately of the LG–HG–LG type. Thus, although the underlying description is formulated in terms of canonical-OAM shell variables, the consequences of the instability are observable in real space through the evolution of the beam profile that can be imaged directly.

Despite the Dzhanibekov analogy as the most vivid dynamical signature, the broader lesson is more structural. In the thin-edge reduction, the boundary field does not merely perturb the state transiently, it is incorporated into the post-edge Hamiltonian and hence into the effective transverse generator itself. The subsequent motion inside the nominally uniform solenoidal region therefore retains a memory of the entry geometry. In this sense, the recurrent mean orbital pseudospin flips are a direct manifestation of \textit{history-dependent} structured-electron transport.

In particular, two otherwise identical electrons, prepared in the same incoming state and propagating in the same nominally uniform magnetic field, can undergo qualitatively different subsequent quasiclassical evolution solely because they entered the field through different magnetic edges.

The mechanism we identified is different from what was previously discussed as it arises in the shell-resolved transport of a twisted charged matter wave, where a nonlinear magnetic entry edge is inherited by the post-edge generator and produces a Dzhanibekov-type instability of the mean orbital pseudospin.

\appendix

%=================================================================================
\section{Field model \label{app:field}}
%=================================================================================

We extend the symmetric $\beta=1/2$ thin-edge field model by introducing a harmonic quartic potential function,
\begin{equation}
\tilde\Phi(\tilde x,\tilde y)=
\frac{B_0}{4}\left[\tilde x^2+ \tilde y^2\right]
+\frac{\kappa_n}{4R_c^2}(\tilde x^4-6\tilde x^2 \tilde y^2+\tilde y^4),
\label{eq:Phi_nonlin}
\end{equation}
with \(\tilde \Delta_\perp \tilde \Phi=B_0\). We then define
\begin{equation}
\tilde A_{\tilde x}=-\theta(\tilde z)\partial_{\tilde y} \tilde \Phi,\qquad
\tilde A_{\tilde y}=\theta(\tilde z)\partial_{\tilde x} \tilde \Phi,\qquad
\tilde A_{\tilde z}=0.
\label{eq:A_nonlin}
\end{equation}
This gauge satisfies \( \tilde \nabla \cdot \tilde{\bm A}=0\) identically and yields
\begin{align}
B_{\tilde x}=-\delta(\tilde z)\partial_{\tilde x} \tilde \Phi,B_{\tilde y}=-\delta(\tilde z)\partial_{\tilde y} \tilde \Phi, 
B_{\tilde z}=\theta(\tilde z)\tilde \Delta_\perp \tilde \Phi.
\end{align}
Thus, we get
\begin{align}
B_{\tilde x}&=
-\delta(\tilde z)\left[\frac{B_0}{2}\tilde x+\frac{\kappa_n}{R_c^2}(\tilde x^3-3\tilde x \tilde y^2)\right],
\nonumber\\
B_{\tilde y}&=
-\delta(\tilde z)\left[\frac{B_0}{2}\tilde y+\frac{\kappa_n}{R_c^2}(\tilde y^3-3\tilde x^2 \tilde y)\right],
\\ \nonumber
B_{\tilde z}&=B_0\theta(\tilde z).
\end{align}
The normal octupole correction modifies only the transverse edge field while preserving the same post-edge longitudinal field and the exact reduction to a two-dimensional transverse Hamiltonian for \(z>0\):
\begin{equation}
\hat {\tilde{H}}_\perp=\frac{(\hat{\tilde{p}}_{\tilde x}-e\tilde{A}_{\tilde x})^2+(\hat{\tilde{p}}_{\tilde y}-e\tilde{A}_{\tilde y})^2}{2k}.
\label{eq:H_nonlin_2D}
\end{equation}
Switching to the normalized coordinates $x=\tilde x/\rho_H$, $y=\tilde y/\rho_H$ and $z=\tilde z/(k \rho_H^2)$ yields
\begin{equation}
\hat H_\perp=\frac{\left[\hat p_x-\operatorname{sgn}(e)A_x\right]^2+\left[\hat p_y-\operatorname{sgn}(e)A_y\right]^2}{2}.
\label{eq:H_nonlin_2DNorm}
\end{equation}
with 
\begin{align}
    \rho_H=\sqrt{\frac{2}{|e|B_0}}
\end{align}
and normalized potentials (for the case of $e<0$ and $z>0$)
\begin{align}
    &\operatorname{sgn}(e) A_x=\left[y+\frac{2\kappa_n}{B_0}\frac{\rho_H^2}{R_c^2}(y^3-3x^2y)\right], \nonumber \\
    &\operatorname{sgn}(e) A_y=-\left[x+\frac{2\kappa_n}{B_0}\frac{\rho_H^2}{R_c^2}(x^3-3xy^2)\right].
\end{align}
\section{Projection to a single orbital sphere \label{app:projection}}

Let
\begin{equation}
\hat{P}_j=\sum_{m_1=-j}^{j}\ket{j,m_1}\bra{j,m_1}
\end{equation}
be the projector onto the fixed Schwinger shell \(\hat N=2j\), where
\begin{equation}
\hat N=\hat{a}_x^\dagger \hat{a}_x+\hat{a}_y^\dagger \hat{a}_y
\end{equation}
and
\begin{equation}
\ket{j,m_1}\equiv \ket{n_x=j+m_1,\ n_y=j-m_1}.
\end{equation}
First, we note that for any scalar function $F$ the following is true 
\begin{align}
\label{eq:comHs}
    [\hat{H}_s, F]=-\frac{i}{2}\left[\nabla F\cdot\hat{\mathbf{p}}_\perp+\hat{\mathbf{p}}_\perp\cdot\nabla F\right]
\end{align}
where
\begin{align}
    \hat{H}_s=\frac{p^2_\perp+\hat{\rho}^2}{2}.
\end{align}
With the help of Eq.\eqref{eq:comHs} and the fact that $\Delta_\perp \Phi_4^{(s)}=0$ one gets
\begin{align}
    \nabla_\perp \Phi_4^{(s)} \cdot\hat{\bm p}_\perp =i[\hat{H}_s,\Phi_4^{(s)}].
\end{align}
Equation above immediately gives 
\begin{align}
    \hat{P}_j\nabla_\perp \Phi_4^{(s)} \cdot\hat{\bm p}_\perp \hat{P}_j=i\hat{P}_j[\hat{H}_s,\Phi_4^{(s)}] \hat{P}_j=0.
\end{align}
Above we used the fact that $\hat{P}_j\hat{H}_s\hat{P}_j=\epsilon_j=const$ as $\hat{H}_s$ is expressed through the Casimir operator and consequently is fixed on the $j$-shell.

Next we consider orbital pseudospin operators in Cartesian Schwinger basis,
\begin{equation}
\hat{\mathcal L}_1\ket{j,m_1}=m_1\ket{j,m_1},
\end{equation}
with
\begin{align}
\hat{\mathcal L}_1&=\frac{\hat{a}_x^\dagger \hat{a}_x-\hat{a}_y^\dagger \hat{a}_y}{2},\nonumber\\
\hat{\mathcal L}_2&=\frac{\hat{a}_x^\dagger \hat{a}_y+\hat{a}_y^\dagger \hat{a}_x}{2},\\
\hat{\mathcal L}_3&=\frac{\hat{a}_x^\dagger \hat{a}_y-\hat{a}_y^\dagger \hat{a}_x}{2i}. \nonumber
\end{align}
The physical orbital angular momentum is
\begin{equation}
\hat L_z=2\hat{\mathcal L}_3,
\end{equation}
so the Laguerre-Gaussian basis used in the main text diagonalizes
\(\hat{\mathcal L}_3\), whereas the Cartesian basis above diagonalizes
\(\hat{\mathcal L}_1\). The shell projector $\hat{P}_j$, however, is basis-independent.

In the symmetric normalized variables (\(\beta=1/2\)),
\begin{equation}
\hat{x}=\frac{\hat{a}_x+\hat{a}_x^\dagger}{\sqrt2},\qquad
\hat{y}=\frac{\hat{a}_y+\hat{a}_y^\dagger}{\sqrt2},
\end{equation}
and the normal octupole quartic part is
\begin{equation}
\hat{V}_4^{(n)}=\frac{\mu_0}{6}(\hat{x}^4-6\hat{x}^2\hat{y}^2+\hat{y}^4).
\label{eq:V4n_app}
\end{equation}
We now project Eq.~\eqref{eq:V4n_app} directly onto the fixed shell.

Using
\begin{align}
&\hat{x}^2=\frac12\qty(\hat{a}_x^2+\hat{a}_x^{\dagger 2}+2\hat{n}_x+1),\nonumber \\
&\hat{y}^2=\frac12\qty(\hat{a}_y^2+\hat{a}_y^{\dagger 2}+2\hat{n}_y+1),
\end{align}
with \(\hat{n}_x=\hat{a}_x^\dagger \hat{a}_x\) and \(\hat{n}_y=\hat{a}_y^\dagger \hat{a}_y\), one finds that only
the \(\Delta N=0\) terms survive under fixed shell projection. Therefore
\begin{align}
\hat{P}_j \hat{x}^4\hat{P}_j
&=
\hat{P}_j\qty(\frac32\,\hat{n}_x^2+\frac32\,\hat{n}_x+\frac34)\hat{P}_j, \nonumber
\\
\hat{P}_j\hat{y}^4\hat{P}_j
&=
\hat{P}_j\qty(\frac{3}{2}\,\hat{n}_y^2+\frac32\,\hat{n}_y+\frac{3}{4})\hat{P}_j,
\\
\hat{P}_j\hat{x}^2\hat{y}^2\hat{P}_j
&=
\hat{P}_j\Biggl[
\frac14(2\hat{n}_x+1)(2\hat{n}_y+1)
\nonumber\\
&+\frac14\qty(\hat{a}_x^{\dagger 2}\hat{a}_y^2+\hat{a}_y^{\dagger 2}\hat{a}_x^2)
\Biggr]\hat{P}_j. \nonumber
\end{align}
Hence
\begin{align}
\label{eq:V4_projected_raw}
&\hat{P}_j\hat{V}_4^{(n)}\hat{P}_j
= \\ \nonumber
&\frac{\mu_0}{6}\hat{P}_j\Biggl[
\frac32(\hat{n}_x^2+\hat{n}_y^2)-6\hat{n}_x\hat{n}_y \nonumber \\ \nonumber
&-\frac32(\hat{n}_x+\hat{n}_y)
-\frac32\qty(\hat{a}_x^{\dagger 2}\hat{a}_y^2+\hat{a}_y^{\dagger 2}\hat{a}_x^2)
\Biggr]\hat{P}_j.
\end{align}

Now rewrite Eq.~\eqref{eq:V4_projected_raw} in terms of generators $\hat{\mathcal L}_i$.

Since
\begin{equation}
\hat{n}_x-\hat{n}_y=2\hat{\mathcal L}_1,
\qquad
\hat{n}_x+\hat{n}_y=\hat N,
\end{equation}
the diagonal part becomes
\begin{align}
&\frac32(\hat{n}_x^2+\hat{n}_y^2)-6\hat{n}_x\hat{n}_y-\frac32(\hat{n}_x+\hat{n}_y)
= \\ \nonumber
&9\hat{\mathcal L}_1^2-\frac34\hat N^2-\frac32\hat N.
\end{align}
On the fixed shell \(\hat N=2j\), this leads to
\begin{align}
&\frac32(\hat{n}_x^2+\hat{n}_y^2)-6\hat{n}_x\hat{n}_y-\frac32(\hat{n}_x+\hat{n}_y)
= \\ \nonumber
&9\hat{\mathcal L}_1^2-3j(j+1).
\end{align}

Next define the ladder operators with respect to the \(\hat{\mathcal L}_1\) axis,
\begin{align}
&\hat{\mathcal L}_+^{(1)}=\hat{\mathcal L}_2+i\hat{\mathcal L}_3=\hat{a}_x^\dagger \hat{a}_y,
\\ \nonumber
&\hat{\mathcal L}_-^{(1)}=\hat{\mathcal L}_2-i\hat{\mathcal L}_3=\hat{a}_y^\dagger \hat{a}_x.
\end{align}
Then
\begin{align}
&\hat{a}_x^{\dagger 2}\hat{a}_y^2+\hat{a}_y^{\dagger 2}\hat{a}_x^2
=
\qty(\hat{\mathcal L}_+^{(1)})^2+\qty(\hat{\mathcal L}_-^{(1)})^2
= \nonumber\\ 
&2\qty(\hat{\mathcal L}_2^2-\hat{\mathcal L}_3^2).
\end{align}
Therefore
\begin{align}
&\hat{P}_j\hat{V}_4^{(n)}\hat{P}_j
= \nonumber\\&\frac{\mu_0}{6}
\hat{P}_j\qty[
9\hat{\mathcal L}_1^2
-3\qty(\hat{\mathcal L}_2^2-\hat{\mathcal L}_3^2)
-3j(j+1)
]\hat{P}_j.
\end{align}
Using the Casimir identity on the shell,
\begin{equation}
\hat{\mathcal L}_1^2+\hat{\mathcal L}_2^2+\hat{\mathcal L}_3^2
=
j(j+1),
\end{equation}
we obtain the exact operator identity
\begin{equation}
\hat{P}_j\hat{V}_4^{(n)}\hat{P}_j
=
\mu_0\qty(\hat{\mathcal L}_1^2-\hat{\mathcal L}_2^2).
\label{eq:V4_final_projection}
\end{equation}

Equation~\eqref{eq:V4_final_projection} is the projected normal octupole
quartic used in the main text.

%++++++++++++++++++++++++++++++++++++++++++++++++++++++++++++++++++
\section{Lewis-Ermakov transport and the quartic normal octupole term \label{app:LEquartic}}
%++++++++++++++++++++++++++++++++++++++++++++++++++++++++++++++++++

After factorizing the exact Larmor rotation, the transverse paraxial dynamics in an axisymmetric magnetic channel is mapped to the unit-frequency oscillator through the metaplectic Lewis-Ermakov transformation \cite{EpovArxiv,Fernandez2003}
\begin{equation}
\hat U_E(z)=\hat M(z)\hat S(z),
\end{equation}
with
\begin{align}
&\hat S(z)=
\exp\!\left[
-\frac{i}{2}\ln b(z)\,
(\hat{\boldsymbol\rho}\!\cdot\!\hat{\mathbf p}_\perp+\hat{\mathbf p}_\perp\!\cdot\!\hat{\boldsymbol\rho})
\right],
\\ \nonumber
&\hat M(z)=
\exp\!\left[
\frac{i}{2}\frac{b'(z)}{b(z)}\hat\rho^2
\right]. 
\end{align}
Here \(b(z)\) satisfies the Ermakov equation
\begin{equation}
b''+\Omega^2(z) b=\frac{1}{b^3}.
\label{eq:app_ermakov}
\end{equation}
In our case $\Omega=0$ for $z<0$ and $\Omega=1$ for the case of $z\geq0$.
For the quartic coordinate operator associated with the normal octupole,
\begin{equation}
\hat Q_4=\hat x^4-6\hat x^2\hat y^2+\hat y^4,
\end{equation}
the transformation is exact and simple. Indeed, \(\hat M\) leaves \(\hat x,\hat y\) invariant, whereas \(\hat S\) acts as a dilation,
\begin{equation}
\hat S\,\hat x\,\hat S^\dagger=b\,\hat x,
\qquad
\hat S\,\hat y\,\hat S^\dagger=b\,\hat y.
\end{equation}
Therefore
\begin{equation}
\hat U_E(z)\,\hat Q_4\,\hat U_E^\dagger(z)=b^4(z)\,\hat Q_4.
\label{eq:Q4_b4}
\end{equation}

Let $\hat{P}_j$ be the static j-shell projector and
\begin{equation}
\hat{P}_j^{E}(z)=\hat U_E(z)\hat{P}_j\,\hat U_E^\dagger(z)
\end{equation}
the transported projector. Using the static shell projection
\begin{equation}
\hat{P}_j\hat Q_4\hat{P}_j=6\qty(\hat{\mathcal L}_1^2-\hat{\mathcal L}_2^2),
\end{equation}
one obtains
\begin{align}
\hat{P}_j^{E}(z)\,\hat U_E\hat Q_4\hat U_E^\dagger&\,\hat{P}_j^{E}(z)
= \\ \nonumber
&6\,b^4(z)\Bigl[
(\hat{\mathcal L}_1^{E}(z))^2-(\hat{\mathcal L}_2^{E}(z))^2
\Bigr],
\label{eq:projected_b4}
\end{align}
where
\begin{equation}
\hat{\mathcal L}_i^{E}(z)=\hat U_E(z)\,\hat{\mathcal L}_i\,\hat U_E^\dagger(z).
\end{equation}

Thus the transported-shell effective Hamiltonian takes the form
\begin{align}
\hat H_{\rm eff}^{E}(z)
&= \nonumber\\
&2\hat{\mathcal L}_3^{E}(z)
+
\mu_0 b^4(z)
\Bigl\{
[\hat{\mathcal L}_1^{E}(z)]^2-[\hat{\mathcal L}_2^{E}(z)]^2
\Bigr\}.
\label{eq:Heff_breathing}
\end{align}

%=================================================================================
\section{Ermakov first integral \label{app:ermakovsol}}
%=================================================================================

For the constant-field case $\Omega=1$ (inside the solenoid $z\geq 0$), Eq.~\eqref{eq:app_ermakov} becomes
\begin{equation}
b''+b=\frac{1}{b^3}.
\label{eq:ermakov1}
\end{equation}
Its first integral is
\begin{equation}
\mathcal E_b=b'^2+b^2+\frac{1}{b^2}=\text{const}.
\label{eq:Eb_def}
\end{equation}
For the initial conditions
\begin{equation}
b(0)=b_0,\qquad b'(0)=0,
\label{eq:b0_ic}
\end{equation}
this constant is
\begin{equation}
\mathcal E_b=b_0^2+b_0^{-2}.
\label{eq:Eb_b0}
\end{equation}
The non-breathing point \(b_0=1\) corresponds to the minimum value
\begin{equation}
\mathcal E_b=2.
\end{equation}
Hence the invariant measure of the breathing amplitude is
\begin{equation}
\mathcal E_b-2=(b_0-b_0^{-1})^2.
\label{eq:Eb_minus2}
\end{equation}

The exact Ermakov solution with \(\Omega=1\) and the initial conditions \eqref{eq:b0_ic} is
\begin{equation}
b^2(z)=b_0^2\cos^2z+b_0^{-2}\sin^2z.
\label{eq:b2_exact}
\end{equation}
Equivalently, in terms of the first integral,
\begin{equation}
b^2(z)=\frac{\mathcal E_b}{2}+\frac{1}{2}\sqrt{\mathcal E_b^2-4}\,\cos 2z.
\label{eq:b2_Eb}
\end{equation}

Squaring,
\begin{equation}
b^4(z)=c_0+c_2\cos 2z+c_4\cos 4z,
\label{eq:b4_harm}
\end{equation}
with
\begin{equation}
c_2=\frac{1}{2}\,\mathcal E_b\sqrt{\mathcal E_b^2-4},
\qquad
c_4=\frac{\mathcal E_b^2-4}{8}.
\label{eq:c2c4}
\end{equation}
Thus \(b_0\neq 1\) is equivalent to \(c_4\neq 0\) and any nontrivial breathing amplitude generates a resonant $4z$ harmonic in the nonlinear coefficient.

%=================================================================================
\section{Linearized pole instability and Floquet exponent \label{app:resonance}}
%=================================================================================

Linearizing the quasiclassical breathing-state dynamics \eqref{eq:ressys} near the north pole,
\begin{equation}
\ell_3\simeq 1,\qquad |\ell_1|,|\ell_2|\ll 1,
\end{equation}
and introducing
\begin{equation}
q(z)=\mu b^4(z),
\qquad
\mu=\frac{\mu_0 j}{2},
\end{equation}
one obtains
\begin{equation}
\ell_1'=-(2+4q)\ell_2,
\qquad
\ell_2'=(2-4q)\ell_1.
\label{eq:lin_firstorder_breathing}
\end{equation}
Eliminating \(\ell_2\) gives
\begin{equation}
\ell_1''-\frac{(1+2q)'}{1+2q}\,\ell_1'
+4(1-4q^2)\ell_1=0.
\label{eq:ell1_secondorder}
\end{equation}
With the substitution
\begin{equation}
\ell_1(z)=\sqrt{1+2q(z)}\,y(z),
\end{equation}
the first-derivative term is removed and one obtains the Hill equation
\begin{equation}
y''+\Xi(z)y=0,
\end{equation}
where
\begin{equation}
\Xi(z)=
4(1-4q^2)
+\frac{q''}{1+2q}
-\frac{3(q')^2}{(1+2q)^2}.
\label{eq:Xi_exact}
\end{equation}

To first order in \(\mu\), using \(q=O(\mu)\), \(q'=O(\mu)\), and \(q''=O(\mu)\), this reduces to
\begin{equation}
y''+\qty[4+q''(z)]y=0+O(\mu^2).
\label{eq:Hill_linear}
\end{equation}
Substituting Eq.~\eqref{eq:b4_harm},
\begin{equation}
b^4(z)=c_0+c_2\cos 2z+c_4\cos 4z,
\end{equation}
yields
\begin{equation}
y''+
\Bigl[
4
-4\mu c_2\cos 2z
-16\mu c_4\cos 4z
\Bigr]y=0+O(\mu^2).
\label{eq:Hill_explicit}
\end{equation}

It is convenient to introduce \(\tau=2z\). Then Eq.~\eqref{eq:Hill_explicit} becomes
\begin{equation}
\frac{d^2y}{d\tau^2}
+
\Bigl[
1
-\mu c_2\cos\tau
-4\mu c_4\cos 2\tau
\Bigr]y=0+O(\mu^2).
\label{eq:Hill_tau}
\end{equation}
The \(\cos 2\tau\) term is in principal parametric resonance with the unperturbed oscillator, whereas the \(\cos\tau\) term is off-resonant at leading order. Since
\begin{equation}
c_4=\frac{\mathcal E_b^2-4}{8},
\end{equation}
any nontrivial breathing amplitude (\(b_0\neq 1\)) produces a resonant modulation because \(c_4>0\).

In the weak-drive limit, standard averaging (or equivalently reduction of Eq.~\eqref{eq:Hill_tau} to Mathieu form) gives the leading Floquet exponent in the \(z\)-variable as
\begin{equation}
\sigma \approx 2\mu |c_4|
=
\frac{\mu}{4}\qty(\mathcal E_b^2-4).
\label{eq:Floquet_Eb}
\end{equation}
Near the non-breathing point \(\mathcal E_b=2\),
\begin{equation}
\sigma \propto \mu(\mathcal E_b-2)
\propto \mu\,|b_0-b_0^{-1}|^2.
\end{equation}
Thus the Ermakov first integral directly controls the strength of the resonant parametric instability.
\bibliographystyle{apsrev4-2}
\bibliography{references}

\end{document}